\documentclass[12pt]{article}
\usepackage{geometry}    
\usepackage{amsmath}
\usepackage{graphicx}
\usepackage{amssymb}
\usepackage{epstopdf}
\usepackage{enumerate}
\usepackage{boxedminipage}
\usepackage{algorithm}
\usepackage{algorithmic}
\usepackage{graphicx}
\usepackage{caption}
\usepackage{subfigure}
\usepackage{multirow}
\numberwithin{equation}{section}
\usepackage{color}
\usepackage{natbib}
\usepackage{rotating}
\usepackage{multirow}
\usepackage{setspace}
\usepackage{titlesec}
\titleformat{\section}[block]{\Large\bfseries\filcenter}{\thesection}{1em}{}
\renewcommand\thesection{\arabic{section}.}

\numberwithin{equation}{section}

\newcommand\T{\rule{0pt}{2.6ex}} 
\newcommand\B{\rule[-1.4ex]{0pt}{0pt}} 
\newenvironment{definition}[1][Definition]{\begin{trivlist}
\item[\hskip \labelsep {\bfseries #1}]}{\end{trivlist}}
\newtheorem{theorem}{Theorem}[section]

\begin{document}
\title{Sequential Bayesian Inference for Dynamic State Space Model Parameters}
\author{Arnab Bhattacharya and Simon Wilson}
\date{}

\maketitle

\section{INTRODUCTION}
\label{sec:intro}

Dynamic state-space models \citep{Durbin:2001aa},  consisting of an unknown state Markov process $X_0, X_1,\ldots$ and noisy observations of that process $Y_1,Y_2,\ldots$ that are conditionally independent, are used in a wide variety of applications e.g.\ wireless networks \citep{Haykin:2004aa}, object tracking \citep{Ristic:2004aa} etc. The model is specified by an initial distribution $p(x_0  |  \theta)$, a transition kernel $p(x_t  |  x_{t-1},\theta)$ and an observation distribution $p(y_t  |  x_t, \theta)$. These distributions are defined in terms of a collection of $K$ static (e.g.\ non-time varying) parameters $\theta=(\theta_1,\ldots,\theta_K)$. The joint model to time $t$ is:
\begin{equation}
p(\mathbf{y}_{1:t},\mathbf{x}_{0:t},\theta) =\left(\prod_{j=1}^t p(y_j  |  x_j, \theta)  p(x_j  |  x_{j-1},\theta) \right) p(x_0 \, | \, \theta) p(\theta),
\label{eq:model}
\end{equation}
where $\mathbf{y}_{1:t} = (y_1,\ldots,y_t)$.

In this paper, we propose a new approach for approximating $p(\theta | \mathbf{y}_{1:t})$. The principal idea is to approximate the posterior on a carefully constructed grid to which points are added or deleted where necessary. Our method is used alongside a large range of state process estimation algorithms making it highly flexible, even for non-linear or non-Gaussian models. We provide evidence corresponding to our claims through a series of examples. 

The literature has tended to focus on computation of the predictive distributions $p(x_t  |  \mathbf{y}_{1:t-1},\theta)$ and $p(y_t  |  \mathbf{y}_{1:t-1},\theta)$, and the filtering distributions $p(x_t  |  \mathbf{y}_{1:t}, \theta)$. Updating of these distributions is by the well-known forward equations \citep{Arulampalam:2002aa}; for example for the prediction of state process we need to compute
\begin{equation}
p(x_t | \mathbf{y}_{1:t-1},\theta) = \int p(x_t | x_{t-1},\theta) p(x_{t-1} | \mathbf{y}_{1:t-1},\theta) \; \mathrm{d}x_{t-1}. \label{eq:prediction1}
\end{equation}
For the linear Gaussian case and assuming known parameters, these computations reduce to the closed form of the Kalman filter \citep{Kalman:1960aa}. Generally, Gaussian forms of the model allows filtering and prediction to be done quickly, exactly and sequentially, without the need to store the data sequence \citep{West:1997aa}. Exact inference has also been achieved for models where the state process is discrete \citep{Cappe:2005aa}.

Extending inference to non-linear and/or non-Gaussian models has proved to be challenging since analytical solutions to the above integrals do not exist. Functional approximation approaches, derived from the Kalman filter, such as the extended Kalman filter \citep{Haykin:2001aa}, unscented Kalman filter \citep{Julier:1997aa} etc have been proposed. Monte Carlo approaches such as the bootstrap filter \citep{Gordon:1993aa} and auxiliary particle filter \citep{Pitt:1999aa} have also been widely used. Attempts have also been made to combine sampling based methods with functional approximations \citep{Merwe:2001ab}.

As regards the static parameter estimation problem, few closed form solutions are available. Different approaches are employed, often involving joint inference of static parameters and the state process \citep{Kantas:2015aa}. We will discuss some of the existing methodologies in the next section.

Section \ref{sec:review} is a review of some of the online static parameter estimation methods, some of which are used later for comparison. Section \ref{sec:principle} outlines the principle of the method. Sections \ref{sec:approx} and \ref{sec:grid} describe the two main issues to be resolved in order to implement the method: approximations to one-step ahead filtering and prediction densities, and updating the grid defining posterior density. A discussion on the theoretical aspects of error accumulation is provided in section \ref{sec:error}  Section \ref{sec:examplesandcomparison} illustrates the method and assesses its performance against alternative approaches. Section \ref{sec:conclude} contains some concluding remarks.

\section{ONLINE PARAMETER INFERENCE REVIEW}
\label{sec:review}

In this paper, interest is in online inference of the the posterior distribution $p(\theta | \mathbf{y}_{1:t})$, along with the filtering and prediction densities associated with the state process (and not on the joint density); a recent review is \citet{Kantas:2015aa}. Several approaches have been proposed in the literature:

\emph{Joint KF/EKF/UKF:}  A common practice in the engineering literature is to add dynamics to static parameters, such as assuming $\theta_t \sim N(\theta_{t-1}, V_t)$ with variance $V_t$ decreasing with $t$, and make inference on $p(x_t,\, \theta_t | \mathbf{y}_{1:t-1})$ at each time point using a single Kalman filter or extensions thereof. One advantage of this is that the state and parameters are typically correlated \textit{a posteriori}, even in linear systems \citep{Haykin:2001aa}; however this is known to suffer from numerical instability issues.

\emph{Dual KF/EKF/UKF:} Dual filters assume that the state and the parameter have separate state space representations, and thus two filters can be run concurrently \citep{Wan:1997aa}. The prediction $p(x_t | \mathbf{y}_{1:t-1},\hat{\theta}_{\text{old}})$ is derived using the parameter mean and  $p(\theta_t | \mathbf{y}_{1:t-1},\hat{x}_{\text{old}})$ is updated using the filtered state mean. Because each posterior only uses the first moment of the other posterior and ignores the variance, these methods are known to produce low variance posteriors.

\emph{Online gradient method:} Sequential optimization of $\log(p(\mathbf{y}_{1:t} \, | \, \theta))$ is also possible. If $\hat{\theta}_{t-1}$ is the estimate after the first $t-1$ observations, it is updated to $\hat{\theta}_t$ after receiving a new data $y_t$ \citep{Poyiadjis:2005aa}. The problem of evaluating the gradient for the whole data  $\mathbf{y}_{1:t}$ has been bypassed in the case of hidden Markov models \citep{LeGland:1997aa}. A problem associated with any gradient method is that it is extremely sensitive to initialization and may converge to a local maximum.

\emph{Online EM:} Online versions of the EM algorithm \citep{Dempster:1977aa}, suitable for a dynamic state space model with unknown model parameters, have been proposed in \citet{Cappe:2011aa}. A major advantage is that it always attempts to maximize the likelihood, allowing methods such as variational inference to be used to estimate the parameters. However this method can also converge to a local maximum.

\emph{Liu and West filter:} This is the most generic among particle filter that performs dual estimation of the state and parameters. Artificial dynamics are introduced for the parameter and subsequently a kernel density estimate of $p(\theta | \mathbf{y}_{1:T})$ is proposed from which $\theta$ can be sampled. Shrinkage is introduced to control for over-dispersion in the kernel density function. A major drawback though is that it requires a significant amount of tuning for quantities such as the kernel bandwidth.

\emph{Storvik's filter:} \cite{Storvik:2002aa} generates particles from the parameter's posterior distribution without assuming any associated random drift. It is further assumed that the posterior distribution of $\theta$ depends on a low-dimensional set of sufficient statistics that can be efficiently updated for each $t$. The choice of this set is the biggest stumbling block for this algorithm.

\emph{Particle learning:} \citet{Carvalho:2010aa} have provided a modified version of Storvik's filter which has proved to be more efficient. Sufficient statistics are derived for the class of conditional dynamic linear models (CDLM), thus providing additional structure to the algorithm. Further, the position of the resampling step in Storvik's algorithm is now interchanged with the propagation step, that allows particle deficiency of the posterior of $\theta$ to be reduced.

For large $t$, particle degeneracy for sampling filters is difficult to avoid, even where artificial drift has been added \citep{Andrieu:2005aa, Kantas:2015aa}.  Online EM and gradient ascent methods do not have this problem. The method in this paper has the same advantage as that of the latter approaches, making it suitable for long sequences of data. It is to be kept in mind that ours is an on-line sequential algorithm and the computational load does not increase over time, unlike other generic methods like $\text{SMC}^2$ \citep{Chopin:2013aa}.

\section{PRINCIPLE}
\label{sec:principle}

The principle of the proposed method is based on two observations.  

The first observation is that many dynamic state space models have a relatively small number of static parameters, so that in principle $p(\theta \, | \, \mathbf{y}_{1:t})$ can be computed and stored on a discrete grid of practical size. This has been noted as a property of many latent models \citep{Rue.:2009aa}.

The second observation is that there are useful identities for parameter estimation in latent models. The one of interest here is a sequential version of the \emph{basic marginal likelihood identity} or BMI \citep{Chib:1995ab},
\begin{equation}
p(\theta | \mathbf{y}_{1:t}) \: \propto \: \left. p(\mathbf{y}_{1:t},\mathbf{x}_{0:t},\theta)\, / \, p(\mathbf{x}_{0:t} |  \mathbf{y}_{1:t},\theta) \right|_{\mathbf{x}_{0:t}=\mathbf{x}^*(\theta)}, \label{eq:inla}
\end{equation}
that is valid for any $\mathbf{x}_{0:t}=\mathbf{x}^*(\theta)$ for which $p(\mathbf{x}_{0:t} |  \mathbf{y}_{1:t},\theta) > 0$ and forms the basis for the integrated nested Laplace approximation (INLA); see \cite{Rue.:2009aa}.

The sequential version of the BMI, which lies at the heart of this approach, is:
\begin{equation}
p(\theta  |  \mathbf{y}_{1:t})  \: \propto  \:  \left. p(\theta  |  \mathbf{y}_{1:t-1}) \: p(y_t  |  x_t, \theta) \: p(x_t  |  \mathbf{y}_{1:t-1},\theta) \, /\, p(x_t  |  \mathbf{y}_{1:t}, \theta) \right|_{x_t = x^*(\theta)}.
\label{eq:sinla2}
\end{equation}
Typically $x^*(\theta) = \arg \max_{x_t} p(x_t  |  \mathbf{y}_{1:t},\theta)$ although the identity is valid for any $x_t$ where $p(x_t  |  \mathbf{y}_{1:t}, \theta) > 0$. 
Equation \ref{eq:sinla2} is useful for sequential estimation because its computation does not grow with $t$, as is the case with Equation \ref{eq:inla}.  Taking Equation \ref{eq:sinla2}, then if prediction and filtering approximations $\tilde{p}(x_t  |  \mathbf{y}_{1:t-1},\theta)$ and  $\tilde{p}(x_t  |  \mathbf{y}_{1:t}, \theta)$ are available, any approximation $\tilde{p}(\theta  |  \mathbf{y}_{1:t-1})$ can be updated:
\begin{equation}
\tilde{p}(\theta | \mathbf{y}_{1:t})  
\: \propto \: \tilde{p}(\theta  |  \mathbf{y}_{1:t-1}) \:  \left. p(y_t  |  x_t, \theta) \: \tilde{p}(x_t  |  \mathbf{y}_{1:t-1},\theta) \, /\, \tilde{p}(x_t  |  \mathbf{y}_{1:t}, \theta) \right|_{x_t = x^*(\theta)}.
\label{eq:sinla_approx}
\end{equation}
Assuming that $\theta$ of low dimension, computing Equation \ref{eq:sinla_approx} on a discrete grid offers the potential for fast sequential estimation. 

Alternatively, the sequential version of Bayes' theorem,
\begin{align}
p(\theta  |  \mathbf{y}_{1:t})  & \propto   p(\theta  |  \mathbf{y}_{1:t-1}) \: p(y_t  |  \mathbf{y}_{1:t-1}, \theta).
\label{eq:sinla1}
\end{align}
and the availability of an approximation $\tilde{p}(y_t  |  \mathbf{y}_{1:t-1}, \theta)$ from any of the filters mentioned, also allows for updating.  Furthermore, using $p(y_t  |  \mathbf{y}_{1:t-1}, \theta)$, one can compute the marginal log likelihood $\log(p(\mathbf{y}_{1:t}))  =   \sum_{j=1}^t\log(p(y_j  \, | \, \mathbf{y}_{1:j-1}))$,
where 
\begin{equation}
p(y_j  |  \mathbf{y}_{1:j-1})  =   \int p(y_j  |  \mathbf{y}_{1:j-1}, \theta) \: p(\theta  |  \mathbf{y}_{1:j})\: \text{d}\theta \: \approx \: \int \tilde{p}(y_j  |  \mathbf{y}_{1:j-1}, \theta) \: \tilde{p}(\theta  |  \mathbf{y}_{1:j})\: \text{d}\theta.
\label{eq:log_likeli2}
\end{equation}
The integration in Equation \ref{eq:log_likeli2} is approximated quickly by a sum over the grid of values of $\theta$.

This suggests the following sequential estimation algorithm when an approximate posterior predictive distribution, or just state prediction and filtering distributions, are available. The process is initiated by choosing a prior $p(\theta)$ and then deterministically selecting a discrete set of points, say $\Theta_0$, to approximately cover the region of high density of the prior, and compute $p(\theta)$ at each of the selected points. The posterior of $\theta$ is updated at each point $\theta_{ti} \in \Theta_t$ by Equation \ref{eq:sinla_approx} or \ref{eq:sinla1}. We would like to reiterate that the objective of this study is not to track the complete trajectory of the state process, but rather to update the posterior parameter and state-filtering density over time.

In this method, $p(\theta \, | \, \mathbf{y}_{1:t})$ may also be approximated by INLA up to some specified time $t=T_{\mbox{\tiny INLA}}$, and from then on the sequential update of Equation \ref{eq:sinla_approx} or \ref{eq:sinla1} is used. This can be particularly helpful because INLA produces a grid $\Theta_t$ that usually provides good support. The value of $T_{\mbox{\tiny INLA}}$ is determined by how quickly the update is required, given that the INLA computation slows down with $t$. Note that $T_{\mbox{\tiny INLA}}$ can be equal to $0$, i.e.\ INLA is not used at all; this means the starting grid is constructed on the prior $p(\theta)$.  

A few issues remain to be addressed in order to implement this algorithm: the form of the approximations $\tilde{p}(y_t  |  \mathbf{y}_{1:t-1}, \theta)$ or that of $\tilde{p}(x_t  |  \mathbf{y}_{1:t-1},\theta)$ and  $\tilde{p}(x_t  |  \mathbf{y}_{1:t}, \theta)$ and how to adapt the grid $\Theta_t$ so that it tracks the support of the posterior over time.  These issues are addressed in the next sections.

\section{POSTERIOR PREDICTIVE AND FILTERING DENSITY APPROXIMATIONS}
\label{sec:approx}
For the Kalman filter, all the distributions required in Equations \ref{eq:sinla2} and \ref{eq:sinla1} are Gaussian, producing exact updates. The means and variances of these Gaussians are sequentially updated. \cite{West:1997aa} is a comprehensive study of models of this type. 

There is a large literature about more general filtering algorithms; see \cite{Doucet:2001aa}. In this paper, we have used functional approximations of the filtering distributions based on extensions to the Kalman filter that incorporate non-linearity and non-Gaussian error while holding on to the basic principle of updating the first two moments. The extended Kalman filter linearizes a non-linear model to create a Gaussian approximation to the filtering and prediction densities. The unscented Kalman filter \citep{Julier:1997aa} also produces Gaussian approximations to the filtering and prediction densities but propagates the means and covariances through the non-linear function. It tends to be more accurate than the extended Kalman filter, more so for strongly non-linear models.

Other finite dimensional suboptimal filters have also been proposed. Some of them are the Gaussian sum filter \citep{Ito:1999aa}, quadrature filters \citep{Davis:2007aa} etc. In our examples, we have used the Kalman filter, the unscented Kalman filter and the quadrature filter. 

\section{UPDATING THE GRID}
\label{sec:grid}
Recall that $\theta$ is of dimension $K$ and assume that there is an initial grid available at time $t=T_{\mbox{\tiny INLA}}$.  If INLA is not used then $T_{\mbox{\tiny INLA}}=0$ and an initial grid $\Theta_0$ is assumed that covers the region of high density of the prior. It is also assumed that the grid is made up of points along a set of basis vectors for $\mathbb{R}^K$, so that it can be written as the Cartesian product
\begin{equation}
\Theta_t = \Theta_{t1} \times \Theta_{t2} \times \cdots \times \Theta_{tK},
\label{eq:theta_cartesian}
\end{equation}
where each $\Theta_{tk} = \{ \theta_{tk1},\ldots,\theta_{tkn_{tk}} \}$ is a set of $n_{tk}$ ordered values for the $k$th component of $\theta$. 

\subsection{Checking the Grid}
For $t > T_{\mbox{\tiny INLA}}$, the grid is checked every $T_{\mbox{\tiny update}}$ observations to see if it is still an appropriate support for $\tilde{p}(\theta \, | \, \mathbf{y}_{1:t})$.

The following simple approach appears to perform well in the examples of the next section. First, the basis that represents the coordinate system for $\theta$ is not changed; it is assumed that by $t = T_{\mbox{\tiny start}}$ there is a basis that will be satisfactory for all future $t$. Second, changes to the grid are based on the univariate marginals of the posterior along the grid coordinates. For the $k$th coordinate, the marginal is:
\begin{align*}
\tilde{p}_k(\theta_{tki}  |  \mathbf{y}_{1:t}) = \sum_{\theta \in \Theta_{t,-k,i}} \tilde{p}\left(\theta \, | \, \mathbf{y}_{1:t} \right) \: \left( \prod_{\substack{l=1\\l \neq k}}^K \delta(\theta_{tl}) \right), \; i=1,\ldots,n_{tk}, 
\end{align*}
where, to give the appropriate marginalizing sum, $\Theta_{t,-k,i}$ is $\Theta_t$ with $\Theta_{tk} = \theta_{tki} \mathbf{1}_{n_k}$, summation is over all coordinates except the $k$th, and $\delta(\theta_{tl})$ is the step size on the grid at $\theta$ along the $l$th coordinate at time $t$.

\begin{figure}
\centering
\includegraphics{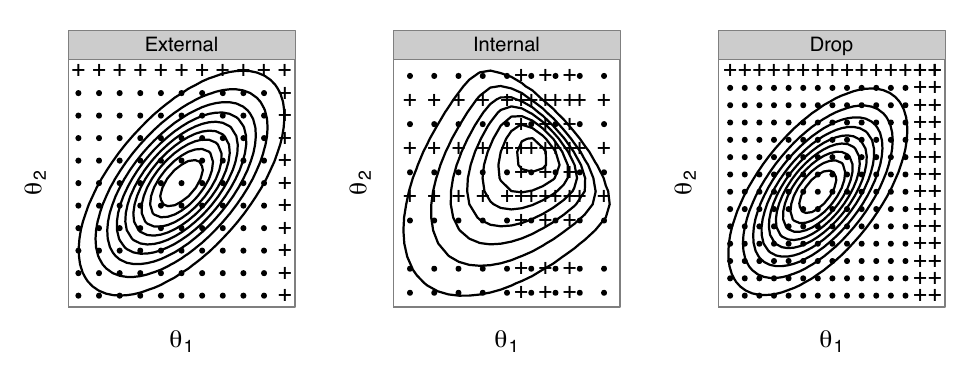}
\caption{Addition and deletion from the grid, where $\bullet$ denotes points that are not changed and $+$ denotes points that are added or deleted. From left to right: external addition, internal addition and external deletion.}
\label{fig:interpol_extrapol_figure}
\end{figure}

Changes are considered when the support of the marginals is not adequately covered by the grid in one of the following situations:
\begin{enumerate}
\item If $\tilde{p}_k(\theta_{tk1}  |  \mathbf{y}_{1:t})$ or $\tilde{p}_k(\theta_{tkn_{tk}}  |  \mathbf{y}_{1:t})$  (i.e.\ the values at each end of the grid support) are greater than a certain proportion, say $\delta_{\mbox{\tiny{ext}}}^{\mbox{\tiny{add}}}$ (chosen to be between 0.15 and 0.3) of the value at the mode of the marginal $\tilde{p}_k(\theta  |  \mathbf{y}_{1:t})$ then an extra point will be added at that end in $\Theta_{t+1,k}$ e.g.\ at $\theta_{nk1} - \delta_k$ and/or $\theta_{nkn_{tk}} + \delta_k$, where $\delta_k$ is a step size along coordinate $k$. Either addition has the effect, through Equation \ref{eq:theta_cartesian}, of adding a $K-1$ dimensional hyperplane to $\Theta_{t+1}$. This is referred to as an external point addition. 
\item Similarly, if either edge value $\tilde{p}_k(\theta_{tk1}  |  \mathbf{y}_{1:t})$ or $\tilde{p}_k(\theta_{tkn_{tk}}  |  \mathbf{y}_{1:t})$ is less than a certain small proportion, say $\delta_{\mbox{\tiny{ext}}}^{\mbox{\tiny{drop}}}$ (chosen to be small, say 0.001) of the marginal mode then either point is removed for $\Theta_{t+1,k}$. Either deletion removes a $K-1$ dimensional hyperplane from $\Theta_{t+1}$. The resulting new edge points in $\Theta_{t+1,k}$ are then also checked and removed if they also satisfy the criterion, allowing more than 1 point to be removed from each end of each $\Theta_{tk}$. Internal grid points are not dropped in this approach. This is referred to as external deletion.
\item If the change in marginal between two existing points, relative to its modal value, 
\[ \left| \frac{\tilde{p}(\theta_{tki} \, | \, \mathbf{y}_{1:t}) - \tilde{p}(\theta_{tk,i-1} \, | \, \mathbf{y}_{1:t})}{\max_{m=1}^{n_{tk}} \tilde{p}(\theta_{tkm} \, |  \, \mathbf{y}_{1:t})} \right|, \] 
is larger than a certain threshold value, say $\delta_{\mbox{\tiny{int}}}^{\mbox{\tiny{add}}}$ (usually chosen between 0.3 and 0.4) then an extra point is added to $\Theta_{t+1,k}$ at $(\theta_{tki} + \theta_{tk,i-1})/2$. This adds a $K-1$ dimensional hyperplane to $\Theta_{t+1}$. This is called internal point addition.
\end{enumerate}
Figure~\ref{fig:interpol_extrapol_figure} shows examples of these additions and deletions. Note that it is possible to increase or decrease the thresholds over time and can be tweaked by the user in advance; for example keeping in mind the fact that the posterior shrinks over time one can assign a lower value to $\delta_{\mbox{\tiny{int}}}^{\mbox{\tiny{add}}}$ later in time such that more points are added to the region of high density.

\subsection{Computing $\mathbf{\tilde{p}(\theta  |  y_{1:t})}$ at added points in $\mathbf{\Theta_{t}}$}

For a set of newly added points to the grid, the most obvious next step is to compute $\tilde{p}(\theta  |  \mathbf{y})$ at any $\theta$ that has been added from $\Theta_{t-1}$.  However, this is computationally infeasible as $t$ gets large and furthermore it would remove the property of the algorithm that it does not require the entire filtering sequence to be rerun from scratch. 

Instead, extrapolation or interpolation methods are used to compute a value of $\tilde{p}(\theta  |  \mathbf{y})$ at any added  external and internal point respectively from the existing values of $\tilde{p}$. Several methods exist e.g.\ kriging \citep{Cressie:1993aa}, thin plate splines \citep{Wahba:1990aa} etc. We use tensor product spline surfaces \citep{Schumaker:2007aa}, more specifically tensor product linear interpolation on the log scale. In our experience so far, this method is both fast and produces good values for new $\tilde{p}(\theta  |  \mathbf{y})$.

Tensor product linear interpolation has also been used to compute the moments of the state estimation filter corresponding to the added external or internal grid points, which in turn provides us with necessary filtering densities such as $p(x_t  |  \mathbf{y}_{1:t},\theta), p(x_t  |  \mathbf{y}_{1:t-1},\theta)$ and $p(y_t  |  \mathbf{y}_{1:t-1},\theta)$. For such points, this simple method works in most cases since we are essentially providing an informative starting value for moments in the Kalman filter framework, even in multivariate settings. Table~\ref{algo1} summarises the complete algorithm.

\begin{table}[!h]
\centering
\caption{The sequential parameter estimation algorithm.}
\label{algo1}
\begin{minipage}[t][7cm][t]{\textwidth}
\hrule
\small
{\tt
\begin{itemize}
\setlength\itemsep{-0.5em}
    \item[] specify $T_{\mbox{\tiny INLA}}$, $T_{\mbox{\tiny update}}$;
    \item[] $t = 0$;
    \item[] if $T_{\mbox{\tiny INLA}} == 0$:
    \item[] \hspace{1cm} specify grid $\Theta_0$ to cover high-density-region of $p(\theta)$;
    \item[] repeat:
        \item[] \hspace{0.4cm} if $t \leq T_{\mbox{\tiny INLA}}$:
            \item[] \hspace{1cm} compute discrete grid $\Theta_t$, i.e.\  $\tilde{p}(\theta  |  \mathbf{y}_{1:t})$ for $\theta \in \Theta_t$ using INLA;
        \item[] \hspace{0.4cm} else:
            \item[] \hspace{1cm} if $(t-T_{\mbox{\tiny INLA}}) \mod T_{\mbox{\tiny update}} == 0$:
                \item[] \hspace{1.4cm} update grid $\Theta_t$ using method of Section \ref{sec:grid};
            \item[] \hspace{1cm} else: $\Theta_t = \Theta_{t-1}$;
            \item[] \hspace{1cm} for each $\theta \in \Theta_t$:
                \item[] \hspace{1.4cm} compute $\tilde{p}(y_t  |  \mathbf{y}_{1:t-1}, \theta)$, or $\tilde{p}(x_t  |  \mathbf{y}_{1:t-1},\theta)$ and  $\tilde{p}(x_t  |  \mathbf{y}_{1:t}, \theta)$, using filtering;
                \item[] \hspace{1.4cm} update $\tilde{p}(\theta  |  \mathbf{y}_{1:t})$ from $\tilde{p}(\theta  |  \mathbf{y}_{1:t-1})$ by Equation \ref{eq:sinla_approx} or \ref{eq:sinla1}.
\end{itemize}
}
\hrule
\end{minipage}
\end{table}


\section{APPROXIMATION ERROR}
\label{sec:error}
Theoretical results about the error in approximating $p(\theta \, | \, y_{1:t})$ using this approach can be divided amongst the various aspects of the approximation:
\begin{enumerate}
\item Error from using a discrete grid;
\item Convergence of the filtering algorithm, including state filtering;
\item Interpolation error in relation to adding grid points.
\end{enumerate}
In this section, existing results around each of these 3 aspects are discussed and a roadmap to proving error bounds is laid out.

It is noted that the unnormalised $\tilde{p}(\theta | \mathbf{y}_{1:t})$ evolves independently at each point in the grid, except for the case of points that are added, where the initial value of $\tilde{p}(\theta | \mathbf{y}_{1:t})$ is an interpolation of neighbouring values. Convergence results for  the polynomial interpolation that is used here are well documented \citep{Kress:1998aa}. For densities with uniformly bounded derivatives, we can ensure that the interpolation is arbitrarily accurate by increasing the order of the polynomial.  However, for the rest of this section, we consider a simpler case where adding/dropping points is not permitted, allowing the third source of error to be ignored and making it sufficient to consider the error in $\tilde{p}(\theta | \mathbf{y}_{1:t})$ at one value $\theta \in \Theta_t$ only, and then separately the consequences of that for errors related to the use of a discrete grid.

For the discretization error, were the approximation to be exact on the grid e.g.\ $\tilde{p}(\theta | \mathbf{y}_{1:t}) \, = \, p(\theta | \mathbf{y}_{1:t}), \forall \theta \in \Theta_t$ then the standard theory of Riemann integration gives us that all integrals of $p(\theta | \mathbf{y}_{1:t})$, at least over compact sets, will be approximated arbitrarily well by finite difference summations of $\tilde{p}(\theta | \mathbf{y}_{1:t})$ as the grid density increases, including over irregular grids.

Therefore the key to results about bounding the error in the approximation of $p(\theta | \mathbf{y}_{1:t})$ by $\tilde{p}(\theta | \mathbf{y}_{1:t})$ lies in the choice of starting values (that come from the prior), error from the use of approximate filters and how that error propagates through to $\tilde{p}(\theta | \mathbf{y}_{1:t})$ by repeated application of Equations \ref{eq:sinla_approx} or \ref{eq:sinla1} at a single grid point. These results are the most challenging of the 3 error sources to discuss. 

At time $t$, given that our algorithm provides the following filtering distribution
\[ \displaystyle \tilde{p}(\theta | y_{1:t}) = C \tilde{p}(\theta) \prod_i \tilde{p}(y_i | y_{1:i-1}, \theta),\]
where $C$ is the accumulated constant of proportionality, the total variation norm between the log of the approximation and the actual data generating process is the following (see Appendix for details):
\[ \displaystyle \left| \log p(\theta | y_{1:t}) - \log \tilde{p}(\theta | y_{1:t}) \right | \le \left | \log p(\theta) - \log \tilde{p}(\theta) \right | + \sum_i \left | \log p(y_i | y_{1:i-1}, \theta) - \log \tilde{p}(y_i | y_{1:i-1}, \theta) \right |.\]
Given that the error is bounded by a sum that is infinite in the limit, it seems that at best it will be possible to show that the error $\left| \log p(\theta | y_{1:t}) - \log \tilde{p}(\theta | y_{1:t}) \right |$ is bounded rather than converging to 0.

Convergence of $\log\tilde{p}(\theta | y_{1:t})$ to $\log p(\theta | y_{1:t})$ can possibly be achieved under the following relatively strong conditions:
\begin{itemize}
\item $\tilde{p}(y_t | y_{1:t-1}, \theta)$ converges to $p(y_t | y_{1:t-1}, \theta)$ as $t \rightarrow \infty$ for each $\theta$.
\item Further, assuming that both $\tilde{p}(y_t | y_{1:t-1}, \theta)$ and $p(y_t | y_{1:t-1}, \theta)$ are bounded away from both 0 and $\infty$ on a compact subset of ${\mathbb R}^d$ in the limit is then sufficient for convergence between $\log\tilde{p}(y_t | y_{1:t-1}, \theta)$ and $\log p(y_t | y_{1:t-1}, \theta)$.
\item Convergence need to be sufficiently fast between $\log\tilde{p}(y_t | y_{1:t-1}, \theta)$ and $\log p(y_t | y_{1:t-1}, \theta)$ such that $\sum_i \left| \log p(y_i | y_{1:i-1}, \theta) - \log \tilde{p}(y_i | y_{1:i-1}, \theta) \right |$ is bounded. For example, geometric convergence of $\log\tilde{p}(y_t | y_{1:t-1}, \theta)$ to $\log p(y_t | y_{1:t-1}, \theta)$ will give a {\it bounded} error 
to $\left| \log p(\theta | y_{1:t}) - \log \tilde{p}(\theta | y_{1:t}) \right |$ by the geometric sum formula.
\end{itemize}

Convergence results in the literature pertain to the state process $x_t$ and hence to differences between $p(x_t \, | \, y_{1:t}, \theta)$ and  $\tilde{p}(x_t \, | \, y_{1:t}, \theta)$. These results can easily be extended to the prediction distribution $\tilde{p}(y_t | y_{1:t-1}, \theta)$.  Focus is on the stability of the filter, so that the total variation norm
\[ \displaystyle \| p(x_t \, | y_{1:t}, \theta) - \tilde{p}(x_t \, | y_{1:t}, \theta) \| \xrightarrow[t \to 0]{} 0.\]

In the simplest setting, convergence results are well researched on:
\begin{itemize}
 \item For the linear filter, properties such as uniform \emph{detectability} and uniform \emph{stabilizability} provide the conditions for convergence \citep{Anderson:1981aa}.
 Since for $\theta \in \Theta_t$, the Kalman filter converges for any $\tilde{p}(x_0)$, convergence for our algorithm would essentially depend only on $\tilde{p}(\theta)$, where $\tilde{p}(\cdot)$ represents the prior chosen by us, while $p(\cdot)$ is the prior in the true data generating process.
\item Finite state space models provide simple settings to study convergence of the filter. Assuming that the actual data generating process arises from $p(\cdot)$, whereas we use $\tilde{p}(\cdot)$ as the prior; one can calculate $p(x_t \, | y_{1:t}, \theta)$ and $\tilde{p}(x_t \, | y_{1:t}, \theta)$ respectively. The total variation between $p(x_t \, | y_{1:t}, \theta)$ and $\tilde{p}(x_t \, | y_{1:t}, \theta)$ is bounded above under ergodic signal, bounded error density, conditions on initial values and certain ``mixing" conditions \citep{Atar:1997ab}. The upper bound is dependent on the Hilbert metric between $\tilde{p}(x_0)$ and $p(x_0)$ and mixing parameters (see Appendix for more details).  For our algorithm, this would mean that convergence of $p(\theta \, | y_{1:t})$ would again be dependent on our choice of the prior $\tilde{p}(\theta)$ along with the aforementioned conditions.
 \end{itemize}
 
Convergence results for non-linear systems are much more difficult to compute; the existing proofs are based on assumptions which are very restrictive, especially in terms of practical application. An illustration of a potential approach in general nonlinear filtering is provided in \citet{Chigansky:2006aa}, where convergence is proved along the same lines as that of finite state space models. 

Most solutions to general state space filtering are sub-optimal, and initial estimate is found to play a very important role. \citet{Kluge:2010aa} and others before them proved that estimation error in EKF is exponentially bounded in mean square under permitted nonlinearity and very conservative bounds for each of initial error and noise covariance matrix. Other filters like UKF or Gaussian filters require tuning of signal covariance matrices to achieve stability; no quantitative bounds have been computed and verification of existing assumptions prove to be exceptionally complex \citep{Xiong:2009aa}. It is extremely difficult to generalise these results to our method, and hence pointwise convergence of $\tilde{p}(\theta | y_{1:t})$ would possibly depend on our choice of prior $\tilde{p}(\theta)$ along with the conditions associated with each filter, as well as those mentioned under general conditions.

A simulation study we have performed substantiates the claim above, vis a vis a linear Gaussian model. Both for state and parameter, priors have been constructed to be extremely narrow and the true starting values for the data generating process are ``very distant" from the region of high density of these priors. Our filter was occasionally found to diverge under these circumstances. An ill-defined state process prior caused more issues than a parameter prior.

\section{EXAMPLES AND COMPARISON WITH ALTERNATIVE APPROACHES}
\label{sec:examplesandcomparison}
 
Our method has been implemented for four real data sets and its performance is compared to offline methods such as INLA, MCMC, iterated filtering  \citep{Ionides:2011aa}and PMCMC \citep{Andrieu:2010aa}, as well as the online method of Liu and West (henceforth referred to as L\&W filter). Computation times have not been compared because of the difficulties in accounting for differences in implementation and code optimization that have been made across all of these methods, even given that they are all written in \texttt{R}. For INLA, the \texttt{R-INLA} package was used (see \emph{www.r-inla.org}), for MCMC in stochastic volatility model the package \texttt{stochvol} was used \citep{Kastner:2016ab} and for iterated filtering, PMCMC and the L\&W filter, the \texttt{R} package \texttt{pomp} was used \citep{King:2010aa}. 

For all the examples the heuristic parameters are set at $\delta_{\mbox{\tiny{ext}}}^{\mbox{\tiny{add}}} = 0.2, \delta_{\mbox{\tiny{ext}}}^{\mbox{\tiny{drop}}} = 0.001$ and $\delta_{\mbox{\tiny{int}}}^{\mbox{\tiny{add}}} = 0.35$. The starting grid has been constructed by placing a regular grid on the region of high density of the prior.
 
\subsection{Nile data} \label{results:subsec_Nile}

The local level model \citep{West:1997aa} is a simple dlm of the form
\begin{align}
y_t &= x_{t-1} + \eta_t, \label{lin:gauss_x_obs}\\
x_t &= x_{t-1} + \epsilon_t, \label{lin:gauss_x_sys}
\end{align}
where $\eta_t \sim N(0, \sigma_{obs}^2)$ and $\epsilon_t \sim N(0,\sigma_{sys}^2)$. Here it is fitted to the well-known Nile data set, being annual measurements of the river Nile at Ashwan from $1871$ to $1970$. 
Paremeter inference is done for $\theta = (\log(\sigma_{obs}^2), \log(\sigma_{sys}^2))$. These data have been elaborately studied by others e.g.\ \citet{Cobb:1978aa}, \citet{Durbin:2001aa} etc. Maximum likelihood estimates are $\hat{\sigma}_{obs}^2 = 15100$ and $\hat{\sigma}_{sys}^2 = 1468$.

The sequential inference algorithm is implemented with a variety of informative and non-informative priors for $\theta$ and $x_0$. The starting grid has been constructed on $40$ points in each dimension. Performance is evaluated qualitatively through trace plots of the posterior median. We would also like to mention at the outset that average runtime was about 3 secs.

Because of a lack of identifiability between $\sigma_{obs}^2$ and $\sigma_{sys}^2$, there is significant sensitivity to the prior and outlier observations. This can cause the approximation to concentrate around a local mode in the posterior.

\begin{figure}
\centering
\includegraphics{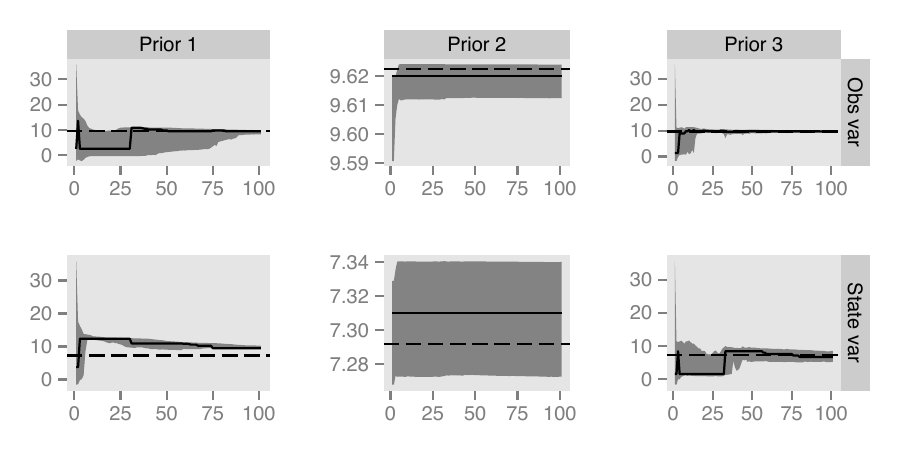}
\caption{Trace plots of the median (solid line), $2.5\%$ and $97.5\%$ posterior quantiles (shaded) of the log variances. The MLE given all the data is a dashed line.}
\label{fig:ex1_Nile_trace_individual}
\end{figure}

This is illustrated in Figure~\ref{fig:ex1_Nile_trace_individual}, where trace plots of parameters with three different prior distributions are given:
\begin{description}
	\item[Prior $1$:] Flat log inverse gamma prior with $\text{shape}=0.1$ and $\text{scale}=1$ for both log variance parameters and an uninformative prior for $x_0$, namely normal with mean $0$ and variance $100$;
	\item[Prior $2$:] An informative log inverse gamma prior for $\text{log}(\sigma_{sys}^2)$ with $\text{shape}=4506$ and $\text{scale}=6760490$, and that for $\text{log}(\sigma_{obs}^2)$ with $\text{shape}=45006$ and $\text{scale}=675015000$. Also an informative normal($1000,100$) for $x_0$;
	\item[Prior $3$:] Flat prior for $\text{log}(\sigma_{sys}^2)$ and $\text{log}(\sigma_{obs}^2)$ (same as Prior $1$) and a prior for $x_0$ similar to Prior $2$ but with larger variance ($10000$).
\end{description}

\begin{figure}
\centering
\includegraphics{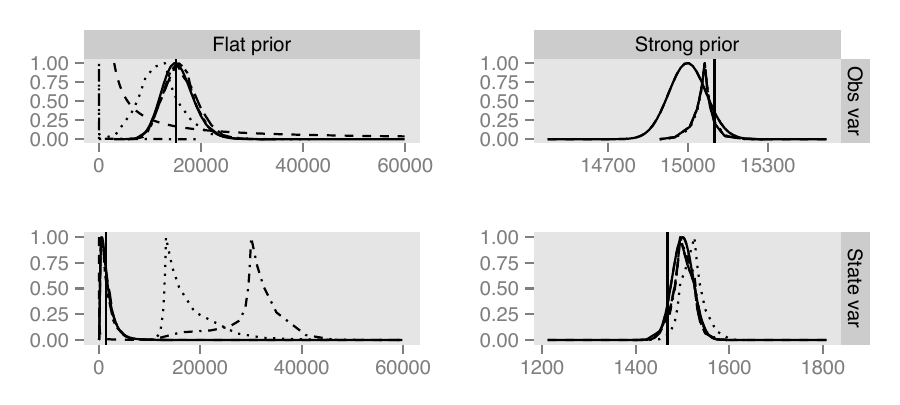}
\caption{Prior density (dashed line) of $\theta$ over-layed on posterior density calculated exactly (solid line) and using our method (all other line types) over various combinations of priors on $x_0$. These densities are scaled to fit. The vertical line is the MLE.}
\label{fig:ex1_Nile_posterior_density}
\end{figure}

It can be seen that in this example, the prior on the state process (or equivalently the starting value for $x_0$) plays the most significant role. The marginal posterior distributions are largely in agreement with the MLE in spite of the prior influence, as is evident in Figure~\ref{fig:ex1_Nile_posterior_density}. It should be mentioned here that when we used INLA to provide the starting grid, our algorithm managed to identify the MLE even when the priors were uninformative for all unknowns (prior $1$).

\begin{figure}
\centering
\includegraphics{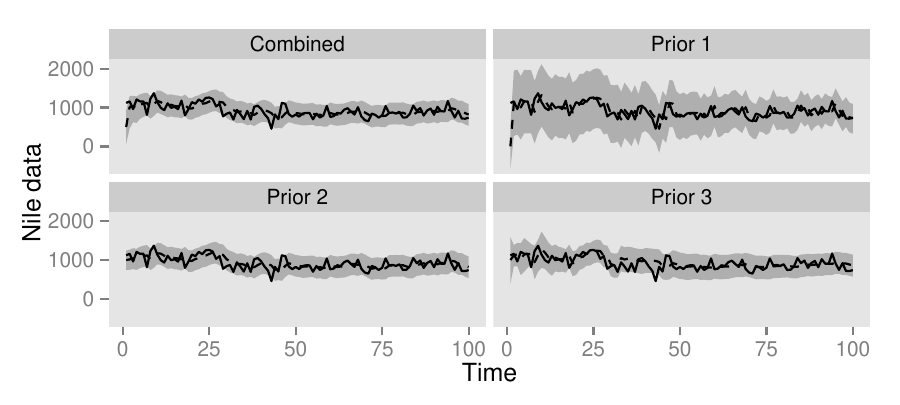}
\caption{The one-step ahead predictive median (dashed line), $2.5\%$ and $97.5\%$ quantiles (shaded) along with data (solid line).}
\label{fig:ex1_pred_trace_all}
\end{figure}

As regards model fit, Figure~\ref{fig:ex1_pred_trace_all} plots the posterior predictions $p(y_{t+1} | \mathbf{y}_{1:t})$ with the data. Predictive performance is good and the effect of the prior is seen in generally tighter prediction intervals where the prior was informative.

It has already been mentioned that the maximum likelihood estimation of the unknown static parameters in this model are $\sigma_{obs}^2 = 15100$ and $\sigma_{sys}^2 = 1468$. 
The L\&W filter was run 50 times for each of the $3$ above mentioned priors. Table~\ref{tab:Nile_table} shows the median of each of the $2.5\%$, $50\%$ and $97.5\%$ posterior quantiles at $t=100$. The methods largely agree under priors 2 and 3, but L\&W fails to track the MLEs in the uninformative prior case (prior 1).
\begin{table}[h]
\begin{center}
\begin{tabular}{ | c | c | c | c | c |}
\hline
 & \multicolumn{2}{ c }{ Sequential BMI\T\B }  & \multicolumn{2} { | c | }{Liu \& West\T\B}\\
\cline{2-5}
 & $\sigma_{state}^2$ \T\B & $\sigma_{obs}^2$ & $\sigma_{state}^2$ & $\sigma_{obs}^2$ \\
\hline
 \multirow{2}{*}{Prior $1$} \T & $15253$ & $10863$ & $76302$ & $131621$\\
  & $(12847, 28340)$ \B & $(3608, 20325)$ & $(15041, 175546)$ & $(26393, 213275)$\\
  \hline
 \multirow{2}{*}{Prior $2$} \T & $1494$ & $15061$ & $1499$ & $15002$\\
  & $(1440, 1541)$ & $(14948, 15123)$ & $(1457, 1548)$ & $(14863, 15137)$\\
  \hline
 \multirow{2}{*}{Prior $3$} \T & $937$ & $15361$ & $1121$ & $15479$\\ 
  & $(181, 4832)$ & $(9530, 21165)$ & $(246, 4991)$ & $(10763, 23214)$\\
 \hline
 \end{tabular}
 \end{center}
 \caption{Comparison of the medians of $2.5\%$, $50\%$ and $97.5\%$ posterior quantiles for the Nile data between the method in this paper and that of L\&W.}
\label{tab:Nile_table}
\end{table}

\subsection{Tokyo rainfall data} \label{results:subsec_Tokyo}
\begin{figure}
\centering
\includegraphics{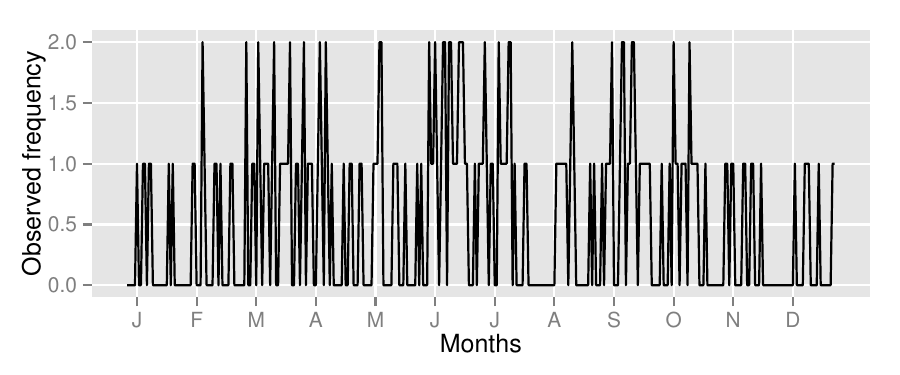}
\caption{Instances of rainfall over $1$ mm in Tokyo for each day from $1983$ to $1984$. The alphabets in the x-axis denote months of the year.}
\label{fig:Tokyo_rainfall_plot}
\end{figure}

Figure~\ref{fig:Tokyo_rainfall_plot} displays the number of instances of rainfall over $1$ mm in Tokyo for each day during the years $1983-1984$ \citep{Kitagawa:1987aa}. For each calendar day $t, \, t = 1, \ldots, 366$ the data is defined as the following:
\begin{align*}
y_t &= 0 \quad \text{if no rain in either year},\\
&= 1 \quad \text{if rain in only } 1 \text{ year and,}\\
&= 2 \quad \text{if rain in both years,}
\end{align*}
with only 1 observation at $t = 60$, corresponding to 29th February $1984$. These data have been analyzed before by several authors, for example \cite{Fahrmeir:2001aa} and \cite{Rue:2005aa}. 

Both \citet{Fahrmeir:2001aa} and \citet{Kitagawa:1987aa} fitted the same model, in which the state process, which is related to the probability $\pi_t = p(\text{rain on day }t)$ through a logit link, is a random walk process. The model is given by,
\begin{align}
y_t &= \begin{cases}\mbox{Bin}(1,\pi_t), & t = 60 \,(\text{February } $29$)\\ \mbox{Bin}(2,\pi_t), & t \ne 60.\end{cases} \label{Tokyo_obs_model}\\
\pi_t &= 1\, /\,(1 + \exp(-x_t)),\label{Tokyo_link}\\
x_t &= x_{t-1} + \epsilon_t \quad \epsilon_t \sim \mathcal{N}(0,\sigma^2).\label{Tokyo_state_rw1}
\end{align}

\citet{Rue:2005aa} use a circular random walk of order $2$ for the state process that cannot be directly analyzed using the ``forward looking" state space approach employed here. Instead a second order difference random walk, also in \citet{Fahrmeir:2000aa}, is used:
\begin{equation}
x_t = 2x_{t-1} + x_{t-2} + \epsilon_t \quad \epsilon_t \sim \mathcal{N}(0,\sigma^2).\label{Tokyo_state_rw2}
\end{equation}

The state variance parameter $\sigma^2$ is the lone parameter in this problem. We have used EKF to compute the estimation in the state process as discussed extensively in \citet{Fahrmeir:1992aa}. A flat prior for the state variance parameter is used and INLA is used for the first $T_{\text{INLA}} = 10$ iterations which is sufficient to get a grid with good posterior support. Average runtime for this data was approximately 30 secs.

\begin{figure}
\centering
\includegraphics{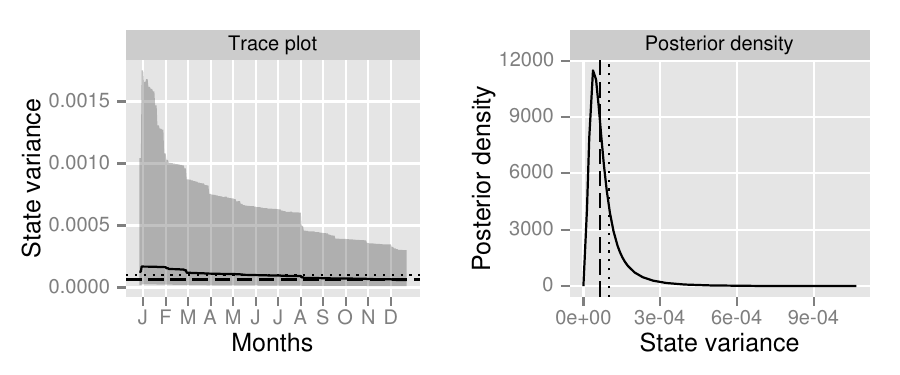}
\caption{The posterior distribution of $\sigma^2$  and comparison to INLA and an MCMC procedure. Left is a summary of $p(\sigma^2 \, | \, \mathbf{y}_{1:t})$: posterior median (solid line), 2.5 and 97.5 percentiles (shaded), posterior mean given all the data from INLA (dashed line) and MCMC (dotted line). Right:  final posterior $p(\sigma^2\, | \, \mathbf{y}_{1:366})$.}
\label{fig:ex2_Tokyo_trace_par}
\end{figure}

Figure~\ref{fig:ex2_Tokyo_trace_par} shows the sequential parameter learning and comparison with 'batch' methods: INLA and an MCMC procedure.
All 3 methods give a similar inference by the time that all the data are observed.

\begin{figure}
\centering
\includegraphics{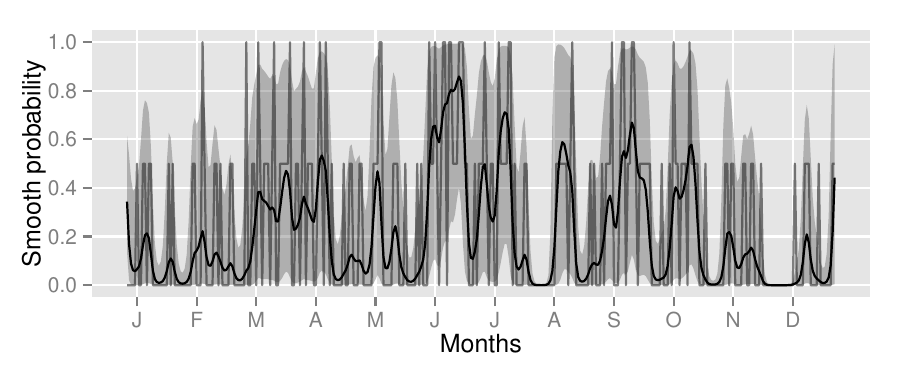}
\caption{Summary of $p(\pi_t\, | \, \mathbf{y}_{1:366}, \hat{\sigma}^2_t)$ where $\hat{\sigma}^2_t$ is the posterior mode: posterior median (solid line) $2.5$ and $97.5$ percentiles (shaded region) and data (solid vertical lines).}
\label{fig:ex2_Tokyo_smooth_predict}
\end{figure}

Figure~\ref{fig:ex2_Tokyo_smooth_predict} plots the smoothing mean $\hat{\pi}_t\, | \, \mathbf{y}_{1:366}, \hat{\sigma}^2_t$ along with bands denoting $2.5\%$ and $97.5\%$ quantiles. The smoothing mean seems to capture the ``rainy patches" in Tokyo quite well and matches with similar inferences made in other works.

\subsection{Stochastic volatility model} \label{results:subsec_stoch_vol}

The stochastic volatility model \citep{Taylor:1982aa} can be written:
\begin{align}
y_t &= \text{exp}\left(x_t\, /\,2 \right) \epsilon_t, \label{SV_obs_model}\\
x_t &= \mu + \phi(x_{t-1} - \mu) + \sigma\eta_t, \label{SV_state_model}\\
x_t &\sim N\left(\mu, \sigma^2\, /\,(1-\phi^2) \right), \notag
\end{align}
where $\epsilon_t$ and $\eta_t$ are independent standard Gaussian variables. The set of unknown static parameters in the model are $\theta = (\mu, \phi, \sigma^2)$: the \emph{level} of log-variance $\mu$, the \emph{persistence} of log-variance $\phi$ and the \emph{volatility} of log-variance $\sigma$.

\begin{figure}
\centering
\includegraphics{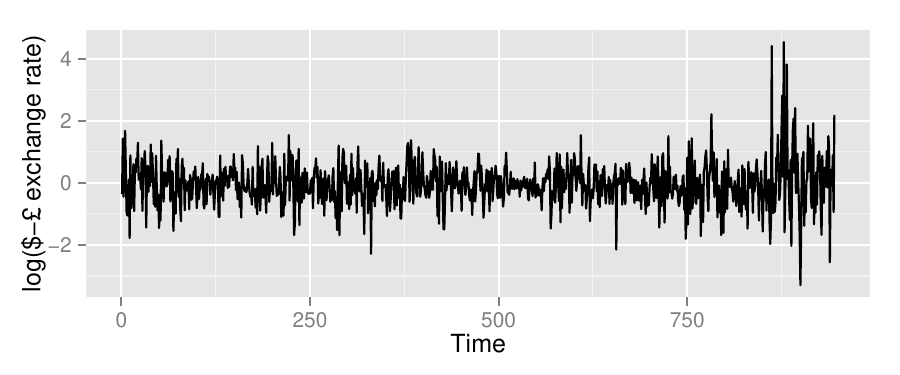}
\caption{Logarithms of daily difference of the dollar-pound exchange rates between October $1^{\text{st}}$ and June $28^{\text{th}}$, $1985$.}
\label{fig:dollar_pound_xchange}
\end{figure}

The daily difference of the dollar-pound exchange rates between October $1^{\text{st}}$ and June $28^{\text{th}}$, $1985$, plotted in Figure~\ref{fig:dollar_pound_xchange}, are analysed with this model.

The algorithm was applied to these data with a set of 32 different priors, corresponding to a mixture of informative and uninformative cases. The starting grids for each prior were 40 for each parameter. 

\begin{figure}
\centering
\includegraphics{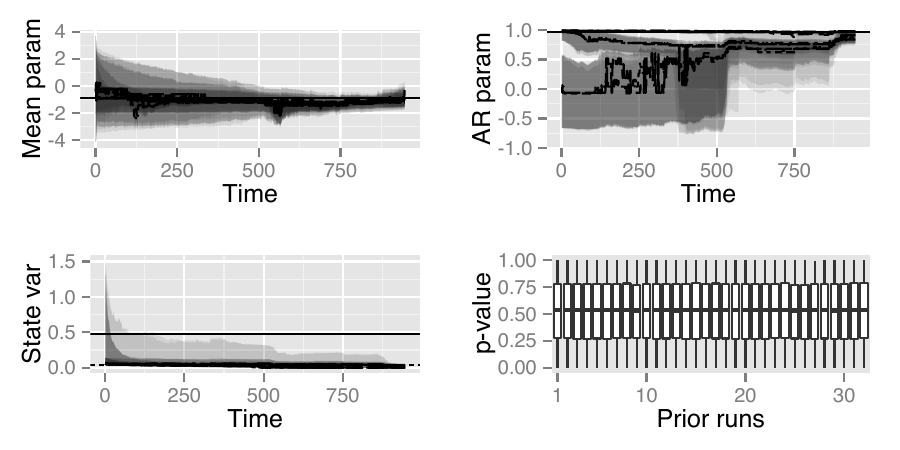}
\caption{Medians of posterior $2.5\%$, $50\%$ and $97.5\%$ quantiles over the 32 prior runs for each of the parameters. The solid and dashed lines correspond to the parameter estimate with all the data from INLA and the MCMC approach. The bottom-right panel plots the Bayesian p-values over various priors on $x_0$.}
\label{fig:stoch_vol_trace_PIT}
\end{figure}

Medians of the posterior quantiles for all the parameters, across the 32 prior specifications, along with $1-$step ahead predictions, are plotted in Figure~\ref{fig:stoch_vol_trace_PIT}, along with the posterior mean from an analysis of all the data using the MCMC approach of \citet{Kastner:2016aa} and INLA. It can be seen that there is good robustness to the prior specification, at least once most of the data are analysed. A much larger estimate of $\sigma^2$ is given by INLA; our method and the MCMC approach are in much better agreement.  The bottom-right panel assesses model fit via box plots of posterior p-values \citep{Gelman:2003aa}, which demonstrates good predictive performance of our method regardless of the prior specifications. The average runtime for our method is approx. 280 sec including calculation of posterior quantiles and p-values.

In addition, Table~\ref{tab:stoch_vol} shows a similar comparison with the method of L\&W. The table summarises the posterior distribution across the 32 priors, showing the median of the posterior median, 2.5\% and 97/5\% quantiles once all the data were analysed. In general, the table shows the difficulties of parameter estimation for this model, with large differences between the approaches.  Our method is in rough agreement with L\&W, but it is also able to provide a reasonable estimate for the AR parameter. INLA showed the largest sensitivity across the different priors that were used. L\&W provide narrower probability bounds than our method.
\begin{table}[h]
\begin{center}
\begin{tabular}{| l |*{2}{ c |} r |}
\hline
Method\T\B              & $\hat{\mu}$ (95\% Intervals) & $\hat{\phi}$ (95\% Intervals) & $\hat{\sigma^2_y}$ (95\% Intervals) \\
\hline
Our method\T & $-0.99$ ($-1.24, -0.53$) & $0.93$ ($0.84, 0.98$) & $0.014$ ($0.002, 0.039$) \\
\hline
Liu \& West\T & $-0.81$ ($-1.29, -0.37$) & $1$ ($1, 1$) & $0.016$ ($0.012, 0.018$) \\
\hline
MCMC\T & $-0.9$ ($-0.96, -0.82$) & $0.968$ ($0.943, 0.975$) & $0.037$ ($0.03, 0.059$) \\
\hline
INLA\T & $-0.9$ ($-0.95, -0.86$) & $0.976$ ($0.956, 0.983$) & $0.49$ ($0.25, 0.72$)\\
\hline
\end{tabular}
\end{center}
\caption{Posterior $2.5\%$, $50\%$ and $97.5\%$ quantiles for each of the parameters from: our approach, L\&W's, MCMC and INLA. }
\label{tab:stoch_vol}
\end{table}

\subsection{Univariate non-stationary growth model} \label{results:subsec_Kitagawa}
The univariate non-stationary growth model (UNGM) is a very challenging model for parameter estimation and has been discussed in numerous occasions in the literature; see \citet{Kitagawa:1996aa, Andrieu:2010aa} for discrete time and \citet{Xu:2009aa} for continuous time version. The model is given by:
\begin{align}
y_t &= x_t^2\, /\,20 + \epsilon_t, \quad \text{ where } \epsilon_t \sim \mathcal{N}(0, \sigma_y^2), \label{Kitagawa_obs_model}\\
x_t &= 0.5\, x_{t-1} + 25\, x_{t-1}\, /\,(1+x_{t-1}^2) + 8\, \text{cos}(1.2\, t) + \eta_t,  \quad \text{ where } \eta_t \sim \mathcal{N}(0, \sigma_x^2),\label{Kitagawa_state_model}\\
x_0 &\sim N(0,1).\notag
\end{align}
The parameters are $\theta = (\sigma_y^2, \sigma_x^2)$. Estimation procedures for parameters for this model have focussed on batch analysis rather than online. As well as being highly nonlinear, both in the state and the observation process, the state process is multimodal, making filtering and prediction difficult, particularly when $x_t^2/2$ is small relative to $\sigma_y$ \citep{Kotecha:2003aa}. 

Data were simulated from this model with $\sigma_y^2 = \sigma_x^2 = 1$. For estimation of the state process, both the augmented UKF \citep{Yuanxin:2005} and quadrature methods were applied, and the latter performed better, at the cost of a longer computation time. Updating the posterior using $p(y_t  |  \mathbf{y}_{1:t-1}, \theta)$ --- that is, the sequential version of Bayes Law --- produced better performance. Another issue is prior sensitivity, which can be considerable because of the nonlinearity and lack of identifiability in the state process. Because of this, to evaluate our method, multiple data sets from this model were generated, while keeping the prior on the state and parameters to be the same in all the runs with the grid to be 40 for both.

\begin{figure}
\centering
\includegraphics{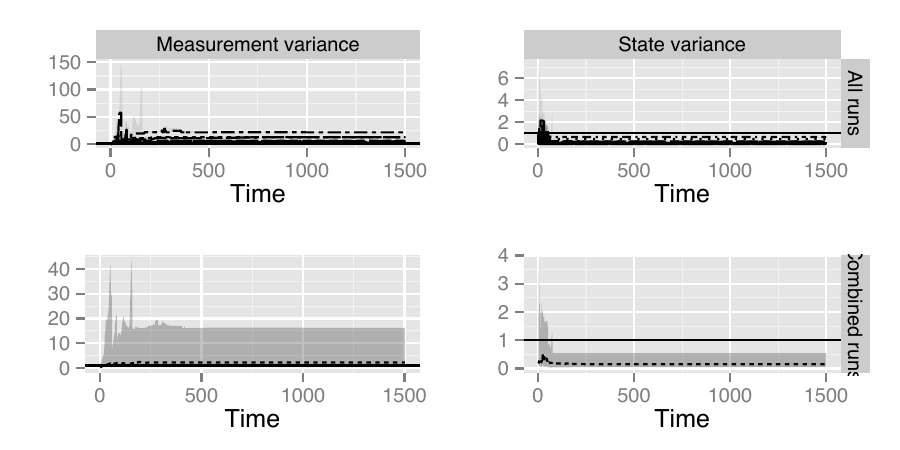}
\caption{Top: posterior modes, $2.5\%$ and $97.5\%$ quantiles for all 24 runs.  Bottom: $2.5\%$, $50\%$ and $97.5\%$ percentiles of the posterior medians from the 24 runs.}
\label{fig:Kitagawa_trace}
\end{figure}

Figure~\ref{fig:Kitagawa_trace} summarises the inference over these $24$ runs which ran over an average time of 850 sec again including computation for posterior quantiles and p-values which slows the process down signifiantly. Note the extremely narrow posterior bounds for each run (top row of the plot), along with consistent under-estimation of the state variance parameter.

\begin{figure}
\centering
\includegraphics{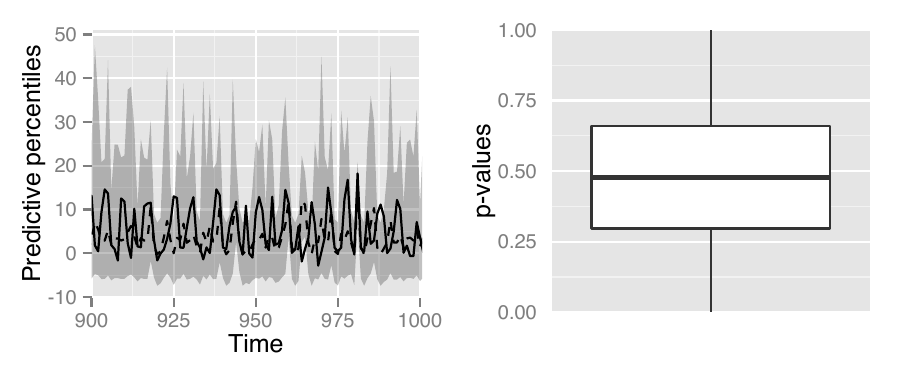}
\caption{Left: snapshot of the $1$-step marginal predictive median (dashed) and probability bounds (band) along with the data (solid) for a single run. Right: box plot of posterior p-values.}
\label{fig:prediction_Kitagawa}
\end{figure}

An explanation for the under-estimation of $\sigma_x^2$ is that while $\sigma_y^2$ has been identified by our method, note the high variability between the runs, mostly indicating over-estimation, an indication of an identifiability problem. In several cases, the posterior distribution diverged considerably from the true value, usually due to the presence of severe nonlinearity. Furthermore, the individual runs were found to have extremely narrow probability bounds, implying that the posterior is concentrated in a narrow region of parameter space. However, Figure~\ref{fig:prediction_Kitagawa}, shows that predictive performance is good (this is for a single run). It is possible to identify the problem faced in identifying the observation when it is close to $0$, i.e.\ when $x_t^2/2$ is small, in the figure. 

Table~\ref{tab:Kitagawa} compares the performance of the L\&W method and two other offline methods: iterated filtering and PMCMC. All the algorithms were provided with a similar set of priors and their posterior medians were recorded. The values provided in the table are the $2.5\%$, $50\%$ and $97.5\%$ percentiles of these medians from multiple runs. L\&W algorithm seem to have estimated the state variance parameter correctly, while our method clearly under-estimated it. However, performance of our algorithm is better for the measurement variance parameter. 
A major disadvantage for the L\&W method is that all the runs ended in a single particle, thus degeneracy is certain, a problem our method does not face. Iterated filtering does contain the true parameter values in its bounds, but at the cost of very large standard errors. PMCMC is the only method which estimates both the parameters correctly, even though it is extremely sensitive to the proposal distribution for the parameters and very easily can get caught in local modes. This goes on to display the extreme complication associated with the UNGM model, because of the nonlinearity in both levels of the state space models. In conclusion one may say that for models with severe nonlinearity and identifiability issues existing online algorithms fail to fit the model well.

\begin{table}[h]
\begin{center}
\begin{tabular}{| l |*{2}{ c |} r |}
\hline
Method\T\B              & $\hat{\sigma^2_x}$ (95\% Intervals) & $\hat{\sigma^2_y}$ (95\% Intervals) \\
\hline
Our method\T & $0.15$ ($0.04, 0.54$) & $2.19$ ($0.27, 16.25$) \\
\hline
Liu \& West\T & $1.19$ ($0.43, 14.6$) & $9.29$ ($0.84, 23.27$) \\
\hline
Integrated filtering\T & $0.47$ ($0.47, 76.93$) & $33.18$ ($0.62, 34.68$) \\
\hline
PMCMC\T & $2.01$ ($0.67, 9.93$) & $3.11$ ($0.52, 5.35$) \\
\hline
\end{tabular}
\end{center}
\caption{The median, with $2.5\%$ to $97.5\%$ percentiles of posterior medians over multiple runs from our approach, L\&W, iterated filtering and PMCMC.}
\label{tab:Kitagawa}
\end{table}

\section{CONCLUSION}
\label{sec:conclude}
A method of sequential static parameter estimation for generalized dynamic state space models has been proposed and compared to multiple alternative approaches, both online and offline.  Our method achieved similar accuracy to the offline methods in almost all the examples provided and usually performed better than the particle filter. Furthermore, it does not suffer from usual scenarios of degeneracy.

Our method has several appealing properties.  It can be applied to models with a non-Gaussian state process, and so is more general than INLA.  When the underlying process is Gaussian, it can make use of INLA for a good initialization.  It is a very flexible framework in that any filtering and prediction algorithm, not just those used here, can be plugged in as the approximations for posterior predictive or filtering distributions. Dynamic grid updating by tensor product splines is fast, simple and seems to work well, at least in the examples described in the paper; many alternative interpolation methods could be implemented once this approach fails to be sufficiently accurate. Finally, the algorithm is trivially parallelizable over the grid; the computation of $\tilde{p}(\theta_j| \mathbf{y}_{1:T})$ at each $\theta_j$ is completely independent.

The principal disadvantage of the approach is that it is restricted to models with a relatively small number of fixed parameters, hence suited for example to inference on hyper-parameters related to the case of hierarchical model in a dynamic setting. Since the number of hyper-parameters in such models are usually much smaller than the number of parameters, which themselves could be time dependent, our algorithm can be fast enough for online Bayesian inference.

Finally, a short discussion on computational complexity of our algorithm is provided. In this paper, our method has not been formally compared to the other online algorithms in terms of computation time, such as L\&W method, because the latter is applied from an R package, which would have more optimized coding than ours. This method has computational complexity $\mathcal{O}(G^p)$ where $G$ is the number of grid points in each dimension and $p = \text{dim}(\theta)$. Because of the recursive nature of the algorithm, the computation is independent of $t$. For each grid point, the complexity also depends on that of the Kalman filter, EKF, UKF or other state estimation methods; total complexity then becomes $\mathcal{O}(G^p \star K)$, where $K$ corresponds to computation in the state estimation process. For the Kalman filter, for example, $K = \mathcal{O}(n^3 + n^2m + nm^2 + m^3)$ where $n$ and $m$ denote the dimensions of the state and observation process respectively. However note that because of the nature of the algorithm, the state estimation steps can be trivially parallelized across points in the grid.

\newpage

\section{Appendix}
\subsection{Convergence of filtering distribution under Hilbert metric approach}
We will first present the general form for a discrete time filter. Let $X = (X_t)$ be a Markov sequence with values in $\mathbb{R}^d$ and transition probability density $p(x_t | x_{t-1}, \theta)$ and initial probability density $p(x_0 | \theta)$. The observation process $Y = (Y_t)$ is an i.i.d sequence conditional of $X$, such that we have $p(y_t | x_t, \theta)$ as the likelihood. We drop the dependence on $\theta$ from future references as they don't play a role in the convergence proofs involving state filtering.

The filtering equation of $X_t$ given $Y_{1:t}$ for the general problem can be written as
\[ \displaystyle \pi_t(x_t) = \frac{p(y_t | x_t)\int_{\mathbb{R}^d} p(x_t | x_{t-1}) \pi_{t-1}(x_{t-1}) dx_{t-1}}{\int_{\mathbb{R}^d} p(y_t | x_t)\int_{\mathbb{R}^d} p(x_t | x_{t-1}) \pi_{t-1}(x_{t-1}) dx_{t-1}\, dx_t}.\]
Even though the filtering equation above is nonlinear, its solution at each time point is obtained by using the linear Zakai equation:
\begin{equation}
\rho_t(x_t) = p(y_t | x_t)\int_{\mathbb{R}^d} p(x_t | x_{t-1}) \pi_{t-1}(x_{t-1}) dx_{t-1}, \text{for }t \ge 0, \label{nonlin:zakai}
\end{equation}
and thus we have
\begin{equation}
\pi_t(x_t) = \frac{\rho_t(x_t)}{\int_{\mathbb{R}^d}\rho_t(u)du}. \label{nonlin:zakai2}
\end{equation}

The \emph{Hilbert projective metric} approach has been used to prove the stability result of the state filter for a finite state space model by \citet{Atar:1997ab} and by \citet{LeGland:2004aa} in particle filters. Let $\mathbb{S} \subseteq \mathbb{R}^d$ be a measurable set and $\mathcal{M}_{+}$ be the space of nonnegative measures on $(\mathbb{S}, \mathbb{B}(\mathbb{S}))$ with the partial order relation $p \preceq q$ if $p(A) \le q(A)$ for any measurable $A \subseteq \mathbb{S}$. The measures $p$ and $q$ are \emph{comparable} if $c_1p \preceq q \preceq c_2p$ for some constants $c_1, c_2 > 0$. 
\begin{definition}
(Hilbert projective distance metric) The Hilbert projective distance is defined as
\[ \displaystyle h(p, q) = \log\frac{\text{sup}_{A, q(A)>0} \, p(A)/q(A)}{\text{inf}_{A, q(A)>0} \, p(A)/q(A)}, \quad p, q \in \mathcal{M}_{+} \text{ are comparable}\]
and $h(p, q) = \infty$ otherwise.
\end{definition}
The following properties of the Hilbert metric are useful:
\begin{enumerate}
\item $h(p, q)$ is a non-negative symmetric function;
\item It satisfies the triangle inequality
\[ \displaystyle h(p, q) \le h(p, r) + h(r, q), \quad p, q, r \in \mathcal{M}_{+};\]
\item $h(p, q) = 0$ iff $p = cq$ for some $c>0$;
\item $h(p, q) = h(c_1p, c_2q)$ for any $p, q \in \mathcal{M}_{+}$ and any scalars $c_1, c_2 > 0$;
\item $\|p - q\| \le \frac{2}{log(3)} h(p, q)$ for any $p, q \in \mathcal{S}^{d-1}$.
\end{enumerate}
Further let $K$ be a linear operator, mapping $\mathcal{M}_{+}$ to itself, then the \emph{Birkhoff contraction coefficient} is defined as
\begin{equation}
\tau(K) = \sup_{0<h(p, q) < \infty} \frac{h(Kp, Kq)}{h(p, q)}. \label{eq:Birkhoff}
\end{equation}
Defining $h$-diameter as $H(K) = \sup_{p, q \in \mathcal{M}_{+}} h(Kp, Kq)$, we have
\begin{equation}
\tau(K) = \tanh \left(\frac{H(K)}{4}\right). \label{eq:h_diam}
\end{equation}
 In the filtering concept, the operator is of a particular integral structure and then if the transition kernel satisfies the ``mixing conditions", i.e.\ if $\exists \text{ constants } \kappa_{\star}$ and $\kappa^{\star}$ s.t.
\[ \displaystyle 0 < \kappa_{\star} \le p(x_t | x_{t-1}) \le \kappa^{\star} < \infty,\]
it can be shown that 
\begin{equation}
\tau(K) \le \frac{\kappa^{\star} - \kappa_{\star}}{\kappa^{\star} + \kappa_{\star}} < 1. \label{eq:tau_mixing}
\end{equation}
Let $\pi = (\pi_t)_{t \ge 0}$ be the filtering posterior that one can get from the data generating process and $\tilde{\pi} = (\tilde{\pi}_t)_{t \ge 0}$ be the applied model, and $p(x_0)$ and $\tilde{p}(x_0)$ be the respective initializations. The condition $p(x_0) << \tilde{p}(x_0)$, i.e. $\tilde{p}_i(x_0) = 0 \implies p_i(x_0) = 0$ is sufficient for $\tilde{p}(x_0)$ to be a valid initialization. Under these assumptions we have the following theorem 
\begin{theorem}
Suppose the signal evolves on a subset $\mathbb{S} \subseteq \mathbb{R}^d$, i.e.\ $p(X_t \notin \mathbb{S}) = 0$ for all $t \ge 0$. Assume that $h(p(x_0), \tilde{p}(x_0)) < \infty$ and there exists a reference measure on $\mathbb{S}$, with respect to which the transition law of the signal has a uniformly positive and bounded density, i.e.\
\[ \displaystyle 0 < \lambda_{\star} \le p(x_t | x_{t-1}) \le \lambda^{\star} < \infty.\]
Then
\[ \displaystyle \| \pi_n - \tilde{\pi}_n\| \le \frac{2}{\log(3)} h(p(x_0), \tilde{p}(x_0)) \left(\frac{\lambda^{\star} - \lambda_{\star}}{\lambda^{\star} + \lambda_{\star}}\right)^n, \quad n \ge 1.\]
\end{theorem}

\subsection{Outline for convergence proof in our algorithm}
Given that our algorithm provides the following filtering distribution:
\begin{align*}
p(\theta | y_{1:t}) &\propto p(\theta | y_{1:t-1}) \, p(y_t | y_{1:t-1}, \theta),\\
&= C_t p(\theta | y_{1:t-1}) \, p(y_t | y_{1:t-1}, \theta),\\
&= C p(\theta) \prod_j p(y_j | y_{1:j-1}, \theta).
\end{align*}
We now have
\[ \displaystyle \log p(\theta | y_{1:t}) = \log p(\theta) + \log C + \sum_j \log p(y_j | y_{1:j-1}, \theta).\]
Assuming $\tilde{p}(\cdot)$ is the approximate filter and $p(\cdot)$ is the true posterior coming from the true data generating process, the total variation norm between the corresponding logs is
\begin{align*}
&\left | \log p(\theta | y_{1:t}) - \log \tilde{p}(\theta | y_{1:t}) \right | \\
& = \left | [\log p(\theta) + \log C + \sum_j \log p(y_j | y_{1:j-1}, \theta)] - [\log \tilde{p}(\theta) + \log \tilde{C} + \sum_j \log \tilde{p}(y_j | y_{1:j-1}, \theta)] \right |,\\
&\le \left | \log p(\theta) - \log \tilde{p}(\theta) \right | + \sum_j \left | \log p(y_j | y_{1:j-1}, \theta) - \log \tilde{p}(y_j | y_{1:j-1}, \theta) \right |.
\end{align*}
We can provide an upper bound to the above inequality provided the conditions mentioned in the main body of the paper hold.

\bibliographystyle{chicago}
\bibliography{masterarnab}

\end{document}